\documentclass[prl,twocolumn,showpacs,preprintnumbers,amsmath,amssymb,superscriptaddress]{revtex4}

\usepackage{graphicx}
\usepackage{dcolumn}
\usepackage{bm}

\begin{document}

\preprint{APS/123-QED}

\title{New electron source concept for single-shot sub-$100\,\rm{fs}$ electron diffraction in the $100\,\rm{keV}$ range}

\author{T. van Oudheusden}
 \affiliation{Department of Applied Physics, Eindhoven University of Technology, P.O Box 513, 5600 MB Eindhoven, The
Netherlands}
\author{E.\,F. de Jong}
 \affiliation{Department of Applied Physics, Eindhoven University of Technology, P.O Box 513, 5600 MB Eindhoven, The
Netherlands}
\author{B.\,J. Siwick}
 \affiliation{Departments of Physics and Chemistry, McGill University, 3600 University St., Montreal, QC. H3A 2T8, Canada}
\author{S.\,B. van der Geer}
 \affiliation{Department of Applied Physics, Eindhoven University of Technology, P.O Box 513, 5600 MB Eindhoven, The
Netherlands}
\author{W.\,P.\,E.\,M. Op 't Root}
 \affiliation{Department of Applied Physics, Eindhoven University of Technology, P.O Box 513, 5600 MB Eindhoven, The
Netherlands}
\author{O.\,J. Luiten}
 \email{O.J.Luiten@tue.nl}
 \affiliation{Department of Applied Physics, Eindhoven University of Technology, P.O Box 513, 5600 MB Eindhoven, The
Netherlands}

\date{Received 21 January 2007; published }

\begin{abstract}
We present a method for producing sub-$100\,\rm{fs}$ electron bunches that are suitable for single-shot ultrafast electron diffraction experiments in the $100\,\rm{keV}$ energy range. A combination of analytical results and state-of-the-art numerical simulations show that it is possible to create $100\,\rm{keV}$, $0.1\,\rm{pC}$, $20\,\rm{fs}$ electron bunches with a spotsize smaller than $500\,\rm{\mu m}$ and a transverse coherence length of $3\,\rm{nm}$, using established technologies in a table-top set-up. The system operates in the space-charge dominated regime to produce energy-correlated bunches that are recompressed by established radio-frequency techniques. With this approach we overcome the Coulomb expansion of the bunch, providing an entirely new ultrafast electron diffraction source concept.
\end{abstract}


\maketitle

The development of a general experimental method for the determination of nonequilibrium structures at the atomic level and femtosecond timescale would provide an extraordinary new window on the microscopic world. Such a method opens up the possibility of making `molecular movies' which show the sequence of atomic configurations between reactant and product during bond-making and bond-breaking events. The observation of such transition states structures has been called one of the holy-grails of chemistry, but is equally important for biology and condensed matter physics \cite{Zewail,Dwyer}.

There are two promising approaches for complete structural characterization on short timescales: Ultrafast X-ray diffraction and ultrafast electron diffraction (UED). These methods use a stroboscopic -but so far multi-shot- approach that can capture the atomic structure of matter at an instant in time. Typically, dynamics are initiated with an ultrashort (pump) light pulse and then -at various delay times- the sample is probed in transmission or reflection with an ultrashort electron \cite{Brad,Ruan} or X-ray pulse \cite{Schotte}. By recording diffraction patterns as a function of the pump-probe delay it is possible to follow various aspects of the real-space atomic configuration of the sample as it evolves. Time resolution is fundamentally limited by the X-ray/electron pulse duration, while structural sensitivity depends on source properties like the beam brightness and the nature of the samples.

Electron diffraction has some unique advantages compared with the X-ray techniques, see e.g. Ref. \cite{Brad-UEO}. However, until recently femtosecond electron diffraction experiments had been considered unlikely. It was thought that the strong Coulombic repulsion (space-charge) present inside of high-charge-density electron bunches produced through photoemission with femtosecond lasers fundamentally limited this technique to picosecond timescales and longer. Several recent developments, however, have resulted in a change of outlook.
Three approaches to circumvent the space-charge problem have been attempted by several groups. The traditional way is to accelerate the bunch to relativistic energies to effectively damp the Coulomb repulsion. Bunches of several hundred femtosecond duration containing high charges (several pC) are routinely available from radio-frequency (RF) photoguns. The application of such a device in a diffraction experiment was recently demonstrated \cite{Dowell}. This is an exciting development; however, energies in the MeV range pose their own difficulties, including the very short de Broglie wavelength $\lambda$ ($\lambda \approx 0.002\,\rm{\AA}$ at $5\,\rm{MeV}$), radiation damage to samples, reduced cross-section for elastic scattering, non-standard detectors and general expense of the technology. Due to these and other considerations, electron crystallographers prefer to work in the $100-300\,\rm{keV}$ range.\\
A second avenue to avoid the space-charge expansion is by reducing the charge of a bunch to approximately one electron, while increasing the repetition frequency to several MHz. According to Ref. \cite{Krausz} by minimizing the jitter of the RF acceleration field the individual electrons arrive at the sample within a time-window of a several fs (possibly even sub-fs). This technique, however, requires that the sample be reproducibly pumped and probed $\sim 10^6$ times to obtain diffraction patterns of sufficient quality.\\
Third, compact electron sources have been engineered to operate in a regime where space-charge broadening of the electron bunch is limited. The current state-of-the-art compact electron gun provides $\sim 400\,\rm{fs}$ electron bunches, containing several thousand electrons per bunch at sub-$100\,\rm{keV}$ energies and with a beam divergence in the $\rm{mrad}$ range \cite{Hebeisen}. This source represents a considerable technical achievement, but is still limited by space-charge effects which place significant restrictions on the applicability of the technique.

The ideal source for single-shot transmission ultrafast electron diffraction (UED) experiments would operate at (several) $100\,\rm{keV}$ energies, providing bunches shorter than $100\,\rm{fs}$, containing $\gtrsim 10^6$ electrons. The transverse coherence length $L_c$ should be at least a few nanometers -or several unit cell dimension- to ensure high-quality diffraction data. None of the electron source concepts presently in use is able to \emph{combine} these bunch requirements. Herein we present a new electron source concept for UED experiments, based on linear space-charge expansion \cite{Jom-PRL} and RF compression strategies, that is able to obtain the ideal parameters presented above with potential well beyond these numbers.

\begin{figure}
    \centering
    \includegraphics[width=\columnwidth]{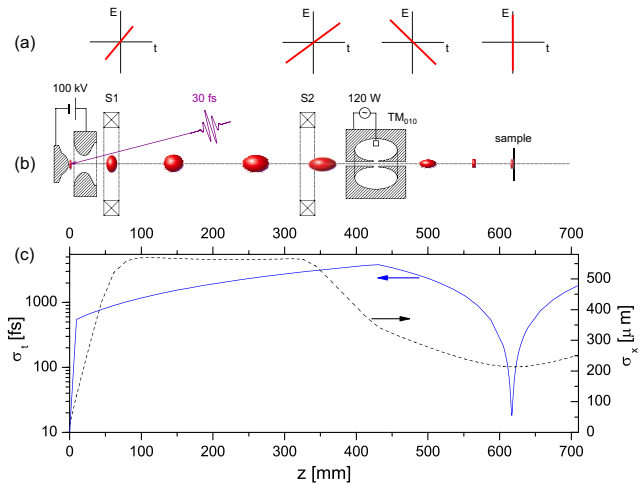}
    \caption{\it (color online.) (a) Schematics of the longitudinal phase-space distribution of the electron bunch at several ``key'' points in the set-up. (b) Schematic of the proposed set-up. The set-up is to scale, the bunches serve only as a guide to the eye. (c) RMS bunch duration (solid line) and RMS bunch radius (dashed line) as function of position.}
    \label{fig:setup}
\end{figure}

Before going into the details of the proposed set-up, it is instructive to first discuss single-shot UED in terms of beam dynamics, because, as we will show, the requirements can \emph{only} be reached by operating close to fundamental space-charge limits. The UED parameter $L_c$ is linked to the conserved beam quantity $\varepsilon_{n}$, the transverse root-mean-squared (RMS) normalized emittance \cite{emittance, Piot}, as follows:

\begin{equation}\label{Lc}
    L_{c} \equiv \frac{\lambda}{2 \pi \sigma_{\theta}} \leq \frac{\hbar}{mc}\frac{\sigma_x}{\varepsilon_{n}},
\end{equation}

\noindent where $\sigma_{\theta}$ is the RMS angular spread, $\sigma_{x}$ the RMS bunch radius, $\hbar$ Planck's constant, $m$ the electron mass, and $c$ the speed of light in vacuum. Requiring $L_c \geq 4\,\rm{nm}$ and $\sigma_x \leq 0.2\,\rm{mm}$ at the sample it follows from Eq. (\ref{Lc}) that $\varepsilon_{n} \leq 0.02\,\rm{mm\,mrad}$. Such low-emittance, ultrashort, highly charged bunches can only be created by pulsed photo-emission \cite{Piot}. The initial emittance for pulsed photo-emission from metal cathodes is $\varepsilon_{n,i} = 8 \times 10^{-3} \sigma_x$ \cite{Piot}, so that a maximum initial RMS radius of $25\,\rm{\mu m}$ at the photocathode is needed. Extracting $0.1\,\rm{pC}$ from such a small spot leads to image-charge and space-charge fields of the order of $1\,\rm{MV/m}$, and therefore requires the acceleration field to be substantially higher (about $10\,\rm{MV/m}$).

During acceleration and the subsequent propagation the bunch expands to millimeter sizes within a nanosecond due to space-charge forces. To be able to compress the bunch, both transversely \emph{and} longitudinally, to the required dimension while conserving its emittance it is necessary that this rapid expansion is reversible; i.e. the space-charge fields inside the bunch must be nearly linear. This is precisely the case for a homogeneously charged ellipsoidal bunch, which has linear internal space-charge fields \cite{Kellogg}. Such a bunch can be created in practice with a ``half-circle'' radial laser profile \cite{Jom-PRL}.
The expansion in the transverse direction can be reversed by (linear) charged-particle optics, for example magnetic solenoid lenses, and in the longitudinal direction by, for example, (linear) RF compression \cite{Bas}. We propose a set-up, consisting of a DC gun, two solenoidal magnetic lenses, and an RF cavity, whose performance we have investigated by detailed particle tracking simulations.

The proposed set-up is shown in Fig. \ref{fig:setup}(b). Electrons are liberated from a metal photocathode by an ultrashort laser pulse and accelerated through a diode structure to an energy of $100\,\rm{keV}$. By applying a DC voltage of $100\,\rm{kV}$ between the cathode and the anode an acceleration field of $10\,\rm{MV/m}$ is obtained. Because of the linear space-charge fields the photoemitted bunch will evolve such that its phase-space distribution becomes linearly chirped with higher energy electrons towards the front and lower energy electrons towards the back. The oscillating electric field in the $\rm{TM_{010}}$ mode in the RF cavity either accelerates or decelerates electrons passing through along the axis, depending on the RF phase. By injecting a bunch just before the field goes through zero, the front electrons are decelerated and the back electrons are accelerated. In such a way the energy-correlation in the bunch can be reversed. Fig. \ref{fig:setup}(a) shows the longitudinal phase-space distribution of the bunch at several ``key'' points in the set-up. During the space-charge-induced expansion the bunch develops a linear energy-position correlation. This correlation is then rotated by the RF cavity leading to ballistic compression in the post-RF-cavity region. An energy difference $\Delta U = 2\,\rm{keV}$ between the most outward electrons is required for ballistic compression of the bunch to $100\,\rm{fs}$, which can be shown by potential energy considerations for a $100\,\rm{keV}$, $0.1\,\rm{pC}$ ellipsoidal bunch of $200\,\rm{\mu m}$ radius. The maximum energy difference that can be introduced by the RF cavity between the most outward electrons of a bunch with duration $\tau$ is given by $\Delta U = e E_0 \omega \tau d$, with $e$ the elementary charge, $E_0$ the RF field amplitude, $\omega$ the frequency of the RF field, and $d$ the cavity length. The required energy difference of $2\,\rm{keV}$ can thus be obtained with an RF field with amplitude $E_0 \approx 3.5\,\rm{MV/m}$, in a cavity with resonant frequency $f = 2 \pi \omega = 3\,\rm{GHz}$ and a length $d = 1\,\rm{cm}$. With the \textsc{superfish} code \cite{Superfish} we have designed an efficient cavity which only requires $120\,\rm{W}$ input power to obtain these fields. Such power can easily be delivered by commercially available solid state RF amplifiers, so klystrons are not required.

The set-up has been designed and optimized with the aid of the General Particle Tracing (\textsc{gpt}) code \cite{GPT}. The bunch charge of 0.1 pC allows us to model the bunch such that each sample particle represents a single electron.

The external fields of both the DC accelerator and the RF-cavity have been calculated with the \textsc{superfish} set of codes \cite{Superfish} with $10\,\rm{\mu m}$ precision. The solenoids are modeled by a 4th order off-axis Taylor expansion from the analytical expression for the on-axis field. The effect of space-charge is accounted for by a Particle In Cell (PIC) method based on a 3-dimensional anisotropic multigrid Poisson solver, tailor made for bunches with extreme aspect ratio's \cite{Bas-SC,Gisela-SC}. Image charges are taken into account by a Dirichlet boundary condition at the cathode.\\
The ideal initial half-circle electron density profile is approximated by a Gaussian transverse profile truncated at a radius of $50\,\rm{\mu m}$ corresponding to the one-sigma point. This profile is experimentally much more easy to realize and turns out to be sufficient. To simulate the photoemission process \textsc{gpt} creates a Gaussian longitudinal charge density profile with a full-width-at-half-maximum (FWHM) duration of $30\,\rm{fs}$. An initial isotropic $0.4\,\rm{eV}$ momentum distribution is used to model the initial emittance.

The RF-phase of the cavity must be tuned to minimize non-linear effects in the longitudinal compression. The optimized phase is a slight deceleration: 11 degrees off the zero-crossing. To compensate for this slight RF-deceleration the voltage of the DC accelerator has been raised from the nominal value of $100\,\rm{kV}$ to $120\,\rm{kV}$ to ensure we have at least $100\,\rm{keV}$ kinetic energy at the sample. Solenoid $S1$ is located at $z = 50\,\rm{mm}$, and produces an on-axis field of $0.05\,\rm{T}$ to collimate the beam. The amplitude of the cavity field is $E_0 = 4\,\rm{MV/m}$, in agreement with the analytical estimate, resulting in less than 1\% relative energy spread at the sample. The optimized position of the RF-cavity, at $z = 430\,\rm{mm}$, is a trade-off between desired longitudinal space-charge expansion to a few ps before injection and unavoidable accumulation of non-linear effects. The position and on-axis field strength of solenoid $S2$, $334\,\rm{mm}$ and $0.03\,\rm{T}$ respectively, have been chosen such that the beam waist at the sample has the desired size and coincides with the time-focus.

\begin{figure}
    \centering
    \includegraphics[width=\columnwidth]{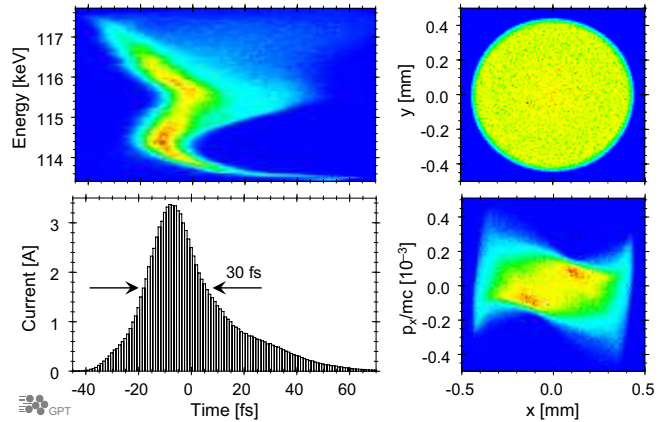}
    \caption{\it (color online.) Longitudinal phase-space distribution, cross-section, current distribution, and transverse phase-space distribution of the electron bunch at the sample.}
    \label{fig:atfocus}
\end{figure}

The bunch evolution in the optimized set-up is shown in Fig. \ref{fig:setup}(c). The transverse beam-size is mainly determined by the two solenoids, but there is also a slightly defocusing effect of the RF cavity. Longitudinally the bunch expands rapidly to several ps due to space-charge forces, to be recompressed by the RF-cavity to below $30\,\rm{fs}$. The current distribution at the sample is shown in Fig. \ref{fig:atfocus}, together with the longitudinal phase-space distribution, the cross-section, and the transverse phase-space distribution. At the sample the $0.1\,\rm{pC}$ bunches are characterized by an RMS duration $\sigma_t = 20\,\rm{fs}$, an RMS radius $\sigma_x = 0.2\,\rm{mm}$, a transverse coherence length $L_c = 3\,\rm{nm}$, an average energy $E = 116\,\rm{keV}$, and relative RMS energy spread $< 1\%$. Of all bunch parameters only the bunch duration is strongly dependent on the longitudinal position: over a range of $5\,\rm{mm}$ around the target position, i.e. $z = (617 \pm 2.5)\,\rm{mm}$, the RMS bunch duration varies between $20\,\rm{fs}$ and $50\,\rm{fs}$, while the other parameters do not change significantly. It is clear that with the presented set-up we are able to create bunches that fulfill \emph{all} the requirements for single-shot UED.

For pump-probe experiments the arrival-time jitter should be less than the bunch duration, requiring a voltage stability of $10^{-6}$ for the power supply of the accelerator. This constraint is also more than sufficient for stable injection on the proper phase of the RF cavity. Such voltage supplies are commercially available. A second requirement is that the laser pulse is synchronized to the RF phase, also with an accuracy of less than the bunch duration. We have developed a synchronization system that fulfills this condition \cite{Fred}.
Because of its direct relationship to the coherence length, see Eq. (\ref{Lc}), the initial spotsize is an important experimental parameter. Simulations, however, show that a deviation of 10\% in spotsize decreases the coherence length by $0.2\,\rm{nm}$ as expected, while the bunch radius and length at the sample do not change significantly.

In summary, we have presented a new robust femtosecond electron source concept that makes use of space-charge driven expansion to produce the energy-correlated bunches required for radio-frequency compression strategies. This method does not try to circumvent the space-charge problem, but instead takes advantage of these dynamics through transverse shaping of a femtosecond laser pulse to ensure the bunch expands in a reversible way \cite{Jom-PRL}. Using this reversibility we propose 6-dimensional phase-space imaging of the electron bunch, with transverse imaging accomplished by regular solenoid lenses and longitudinal imaging by RF bunch compression. Our \textsc{gpt} simulations show that it is possible to create $0.1\,\rm{pC}$, sub-$100\,\rm{fs}$ bunches at sub-relativistic energies in realistic accelerating and focusing fields. We have designed a compact set-up to create electron bunches that are suitable for single-shot, ultrafast electron diffraction experiments. With these bunches it will truly be possible for chemists, physicists, and biologists to study atomic level structural dynamics on the sub-$100\,\rm{fs}$ timescale.

\bibliography{UEDbib}

\end{document}